\documentclass[10pt,twocolumn]{article}
\usepackage{epsfig}\usepackage{amstex}
\begin{document}
\title{Orthogonality catastrophe in a composite fermion liquid}
\author{Darren J. T. Leonard, T. Portengen, V. Nikos Nicopoulos and Neil
F. Johnson.\\
Department of Physics,\\ Clarendon Laboratory,\\ Oxford University,\\
Oxford OX1 3PU,\\ England.}
\maketitle
\begin{abstract}
We discuss the emergence of an orthogonality catastrophe in the response of
a composite fermion liquid as the
filling factor \(\nu\) approaches \(1/2m\), where \(m=1,2,3...\). A
tunneling experiment is proposed in which 
dramatic changes in the 
I-V characteristic should be observable as \(\nu\) is varied. Explicit I-V characteristics
calculated within the so-called Modified Random Phase Approximation, are
provided
for \(\nu=1/3\rightarrow 1/2\). 
\end{abstract}
\section*{Article}
Composite fermion theory~\cite{HLR,STERN,SIMON} has been remarkably
successful in explaining the
fractional quantum Hall effect
in terms of the integer quantum Hall effect of a composite fermion
metal~\cite{DU,SAW}.
Within this theory, the ground state of a two-dimensional electron gas
(2DEG)
at even-denominator filling factors \(\nu=1/2m,\ m=1,2,3\ldots,\) is a
compressible Fermi liquid 
containing composite fermions which experience zero effective magnetic field
\(B^\star\) ~\cite{HLR,STERN,SIMON}. 
Various recent experiments on systems near \(\nu=1/2\) 
have indeed been interpreted in terms of a Fermi sea of composite fermions~\cite{DU,SAW}.
Further confirmation of the details of the theory as \(\nu\rightarrow 1/2m\) is still however
desirable. 
In this preprint, we propose an experiment to probe the 
emergence of the Fermi surface in a composite fermion liquid as
\(\nu\rightarrow1/2m\). 
Mono-energetic electrons are allowed to tunnel into a quantum dot placed
close to the 
2DEG. We find the signature of the presence of a Fermi surface at \(\nu=1/2m\)
to be a dramatic `orthogonality catastrophe' in the tunneling
current (I) as a function of
the gate voltage (V) of the dot. 
For \(\nu\neq 1/2m\), strong oscillatory signatures 
in the I-V characteristic are predicted -- 
these signatures are \(\nu\)-dependent and can be used to deduce the
composite fermion effective 
mass. 
In addition, the spectrum for odd-denominator
fractions will yield valuable information about the `gapfull' excitations of
these states. 
\par
Our proposed experiment 
is analogous to inverse photoemission spectroscopy (IPS)\cite{MAHAN,IPS}
experiments on ordinary metals, 
with the atomic core level replaced by a quantum dot. 
In an IPS experiment, a free electron falls into a hole in an atomic core
state
thereby emitting a photon of energy \(\omega_0\). The
intensity of the emitted photon is identically zero at the threshold energy
\(\omega_0\). 
This is a consequence of the so-called `orthogonality catastrophe' (OC); the
transition involved 
is forbidden because the initial and final states are
orthogonal~\cite{MAHAN,IPS,MATVEEV,SHAM,NOZ}.
\par
\begin{figure}
\centering\epsfig{file=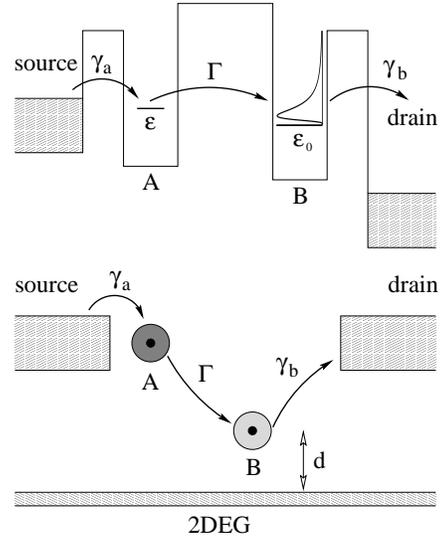,width=0.7\linewidth,angle=0}
\caption{Schematic diagram of the junction. A gate voltage \(V\) is applied to dot B
to alter the 
energy of dot B with respect to A. 
The source-drain voltage is kept fixed. The plane of the 2DEG is
perpendicular to the page.}
\label{fig:JUNCTION}
\end{figure}

Figure~\ref{fig:JUNCTION} provides a schematic illustration of our proposed
experiment.
Quantum dot A acts as an electron monochromator, because only 
source electrons with energy \(\epsilon\) can resonantly tunnel to A. An
electron in A will resonantly
tunnel to quantum dot B only if there are states available with energy
\(\epsilon\). In the absence of the
2DEG the density of states of B is a \(\delta-\)function at energy
\(\epsilon_0\), and tunneling occurs
only when  \(\epsilon=\epsilon_0\). However, the presence of the 2DEG means
that the density of states is 
asymmetrically broadened to higher energies due to the neutral excitations
in the 
2DEG induced by the filled dot -- this implies that tunneling can occur 
if \(\epsilon>\epsilon_0\). The electron can then tunnel 
out into the drain lead to be measured as a current, determined by the
tunneling rates 
\(\gamma_a,\ \Gamma,\text{ and }\gamma_b\).
The current \(I\) is measured as function of the 
gate voltage \(V\) controlling the difference between the 
energy \(\epsilon\) of the injected electron and the energy \(\epsilon_0\)
of dot B.
This resonant tunneling is similar to IPS with a zero energy photon, and the
analog of the IPS
spectrum is the tunneling I-V characteristic; the threshold \(V=V_0\) in
this case is such that 
\(\epsilon_0=\epsilon\). Generally, \(\epsilon-\epsilon_0=e(V-V_0)\), hence
the creation of excitations implies that the spectrum is non-zero for
\(V>V_0\). 
\par 
Once an electron has tunneled into dot B it must reside there for a time
greater than the response time
of the 2DEG. This implies that the electron tunnels out with rate
\(\gamma_b\) less than the desired resolution, which is typically the 
composite fermion Landau-level spacing for 
a filling factor close to \(\nu=1/2\). The other two rates \(\Gamma\text{
and }\gamma_a\) are determined by 
the following simple argument.
The average current can be written as
\begin{equation}
I=\frac{e}{T}
\end{equation}
where \(T\) is the total time taken to tunnel from source to drain. In terms
of the tunneling rates we have
for sequential tunneling
\begin{equation}
T\sim\frac{1}{\gamma_a}+\frac{1}{\Gamma}+\frac{1}{\gamma_b}.
\end{equation}
In order for the tunneling current to reflect only the density of states
of dot B, we choose 
the conditions
\begin{equation}\label{eq:CONDITIONS}
\gamma_a \sim \gamma_b \equiv \gamma \hspace{3em} \Gamma \ll \gamma
\hspace{3em} \gamma \leq \omega_c^\star,
\end{equation} 
in which case the current is given by
\begin{equation}\label{eq:CURRENT}
I=e\Gamma=e\Gamma_0 \gamma \, \text{Re}\,\int_0^\infty
dt\,e^{ie(V-V_0)t/\hbar-F(t)-\gamma t}.
\end{equation}
The time-integral in Eq.~(\ref{eq:CURRENT}) gives the convolution of the
density of states 
of dot B with a Lorentzian of width \(\gamma\) representing the broadening
due to tunneling to the 
drain. In the absence of the 2DEG, the function \(F(t)=0\), and
\(I(V_0)=e\Gamma_0\). 
The bare tunneling rate \(\Gamma_0\) is 
determined solely by the width and height of the barrier between dots A and
B. 
A suitable value consistent with Eq.~(\ref{eq:CONDITIONS}) is
\(\hbar\Gamma_0\approx 6\mu\)eV. 
The density of states of the electron, once it has tunneled to dot B, depends
on the excitation spectrum of 
the 2DEG through the function
\begin{equation}
F(t)=\int_0^\infty d\omega\,
\frac{(1-e^{-i\omega t})}{\omega^2}\,\rho(\omega)
\end{equation}
where
\begin{equation}
\rho(\omega)=\frac{1}{\hbar}\,\sum_{\mathbf{q}}\,|V({\mathbf{q}})|^2\,
\,S({\mathbf{q}},\omega)
\end{equation}
is the density of single pair-excitations of the 2DEG due to the sudden
appearance of an electron in B. 
\(V({\mathbf{q}})\) is the potential experienced by the 2DEG due to the
electron in B~\cite{MAHAN}.
The dynamic structure factor \(S({\mathbf{q}},\omega)\)~\cite{PINES}
contains
information about the excitation spectrum of the 2DEG, and is calculated
using the 
Chern-Simons theory of composite fermions within the modified random phase
approximation (MRPA)~\cite{SIMON}.
We note that in approximating \(F(t)\) as above, we are treating the 
excitations of momentum \({\mathbf{q}}\) and energy \(\omega\) as
independent bosons, which  
is standard in the theory of IPS in ordinary metals. The theory includes all such
bosonic
excitations exactly to all orders in perturbation theory~\cite{MAHAN}. 
\par
The MRPA resolves the conflict of requiring both renormalisation of the
composite fermion
mass and satisfaction of Kohn's theorem~\cite{HLR,STERN,SIMON,PINES}. In the
limit where the electron cyclotron
energy is large compared with the Coulomb energy, the composite fermion mass
is expected to scale
as the square-root of the magnetic field. Using the RPA equations with this
renormalised
mass leads to satisfying neither Kohn's theorem nor the f-sum rule; the MRPA
repairs this within 
Fermi-liquid theory by adding a Landau interaction term. 
In our calculations we use a renormalised composite fermion effective mass
which
scales as the square-root of the magnetic field such that
\begin{equation}\label{eq:MASS}
m^\star_{\text{CF}}=\frac{4\pi\epsilon_0\epsilon_r\hbar^2}{0.3e^2
l_{\text{c}}},
\end{equation}
where the magnetic length is \(l_{\text{c}}\) and the dielectric constant
\(\epsilon_r=13\)~\cite{HLR,STERN,SIMON}. 
\par
There are three essential features to be incorporated in the 
experimental design (see Fig.~\ref{fig:JUNCTION}). First, the probe dot B must be
separated
 from the plane of the 2DEG
by a barrier which is sufficiently high to prevent tunneling between B and
the 2DEG. 
However, B must be as close
to the 2DEG as possible because the Fourier transform of the potential
experienced by the 2DEG 
due to an electron at B decays exponentially with the separation \(d\).
Second, the source and drain leads 
and dot A must be far enough away from the 2DEG so that the only potential
experienced by the 2DEG 
comes from the electron at B. Third, the levels in the dots must be 
well spaced so that only one level
contributes to the tunneling; two-electron
tunneling will be suppressed because of Coulomb blockade. 
Our calculations are therefore based on a 2DEG with \(10^{15}\text{
electrons m}^{-2}\) placed 
\(d=50\) \AA\ away from a dot with a confinement length of 50 \AA. 
\par
\begin{figure}
\centering\epsfig{file=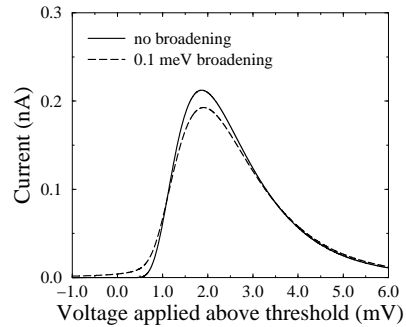,width=0.7\linewidth,angle=0}
\caption{Current at \(\nu=1/2\) as a function of the
voltage applied above threshold,
i.e. \(V-V_0\). Solid curve: spectrum in limit of perfect resolution, i.e. no 
instrumental broadening. Dashed curve: current with a broadening of 0.1
meV.}
\label{fig:CURRENT1}
\end{figure}

Figure~\ref{fig:CURRENT1} shows the I-V characteristic calculated for the
compressible 
state with filling factor \(\nu=1/2\). The solid line does not include
any instrumental
broadening thereby emphasising the suppression at threshold.
The dashed curve shows the current with a Lorentzian broadening of width
\(\gamma=0.1\) meV; 
here there is a small
current at and below the threshold. 
The peaked shape is completely different from the
power-law
which arises for an ordinary metal~\cite{MAHAN,IPS,MATVEEV,SHAM,NOZ}. The OC is
clear, because the threshold current is negligible - the low energy excited
states have a very small overlap
with the initial state and are strongly suppressed, in
complete contrast to those in an ordinary metal. The suppression is a
consequence of the diffusive mode 
which dominates the dynamic structure factor at \(\nu=1/2\) at low energies
and momenta~\cite{HLR}. 
This mode arises from the scattering of the composite fermions by their attached flux
tubes. 
The sudden appearance of the perturbing potential causes density
fluctuations in the 2DEG,
which induce modulations in both the scalar and vector potentials resulting
from the diagonal and off-diagonal terms in the Chern-Simons interaction; 
these potentials further scatter the composite fermions thereby
creating more density fluctuations. The result is that these density fluctuations
are diffusive, i.e. they decay exponentially with time. The excitations do not 
exist on a time scale long enough to screen
the potential. This situation is very different in an ordinary metal,
where the only scattering mechanism available is the Coulomb potential. 
In this case, the low energy excitations have infinite lifetime,
hence they successfully screen the potential in the long-time limit.
\par 
The power-law divergence in ordinary metals arises from a density of pair
excitations \(\rho(\omega)\)
which goes linearly with \(\omega\) for energies small compared with the
Fermi energy~\cite{MAHAN}. 
This is not the 
case for the composite fermion metal, where we find \(\rho(\omega)\) varies
roughly as \(\sqrt{\omega}\). One can derive 
this result analytically using the single-mode approximation employed 
by Platzman, He and Halperin~\cite{PLATZMAN} in their 
study of tunneling between two 2DEGs as a function of the bias voltage. It
should be noted that  
our work differs in two ways from that of Ref.~\cite{PLATZMAN}.
First, and most importantly, their problem involves the \emph{charged}
excitations of the 2DEG 
because their tunneling process removes an electron 
from one 2DEG and places it in the other. They treat the added (removed)
electron as an infinitely massive 
foreign particle inserted into the N-electron
system, thereby reducing the problem to that of the x-ray edge problem. Our
experiment is, by design,
analogous to IPS without further need for approximations because 
the tunneling electron is both distinguishable from the 2DEG and localised.
Our experiment therefore strictly probes the \emph{neutral} excitations of
the 2DEG. 
Second, our work differs in the level of approximation employed for the 
dynamic structure factor and the subsequent determination of the density of
pair excitations.
We use the full MRPA dynamic structure factor, which allows us to calculate the I-V 
characteristic at both odd- and
even-denominator filling factors. By contrast, the single-mode approximation used in 
Ref.~\cite{PLATZMAN} is valid only at even-denominator
filling factors. Applying the single-mode approximation to our experiment (with \(d=0\))
we predict a current of the form
\begin{equation}
I\propto\frac{1}{\sqrt{(V-V_0)^3}}\exp\left(\frac{\alpha}{V_0-V}\right)\hspace{1.5em}V>V_0
\end{equation}
which is in qualitative agreement with the current calculated numerically.
\par
\begin{figure}
\centering\epsfig{file=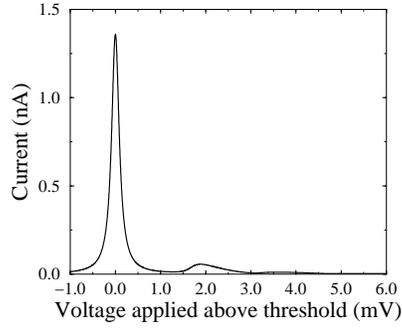,width=0.7\linewidth,angle=0}
\caption{Current calculated for \(\nu=1/3\) as a function of the voltage
applied above threshold, i.e.
\(V-V_0\). Finite broadening as in Fig.~\ref{fig:CURRENT1}.}
\label{fig:CURRENT2}
\end{figure}
\begin{figure}
\centering\epsfig{file=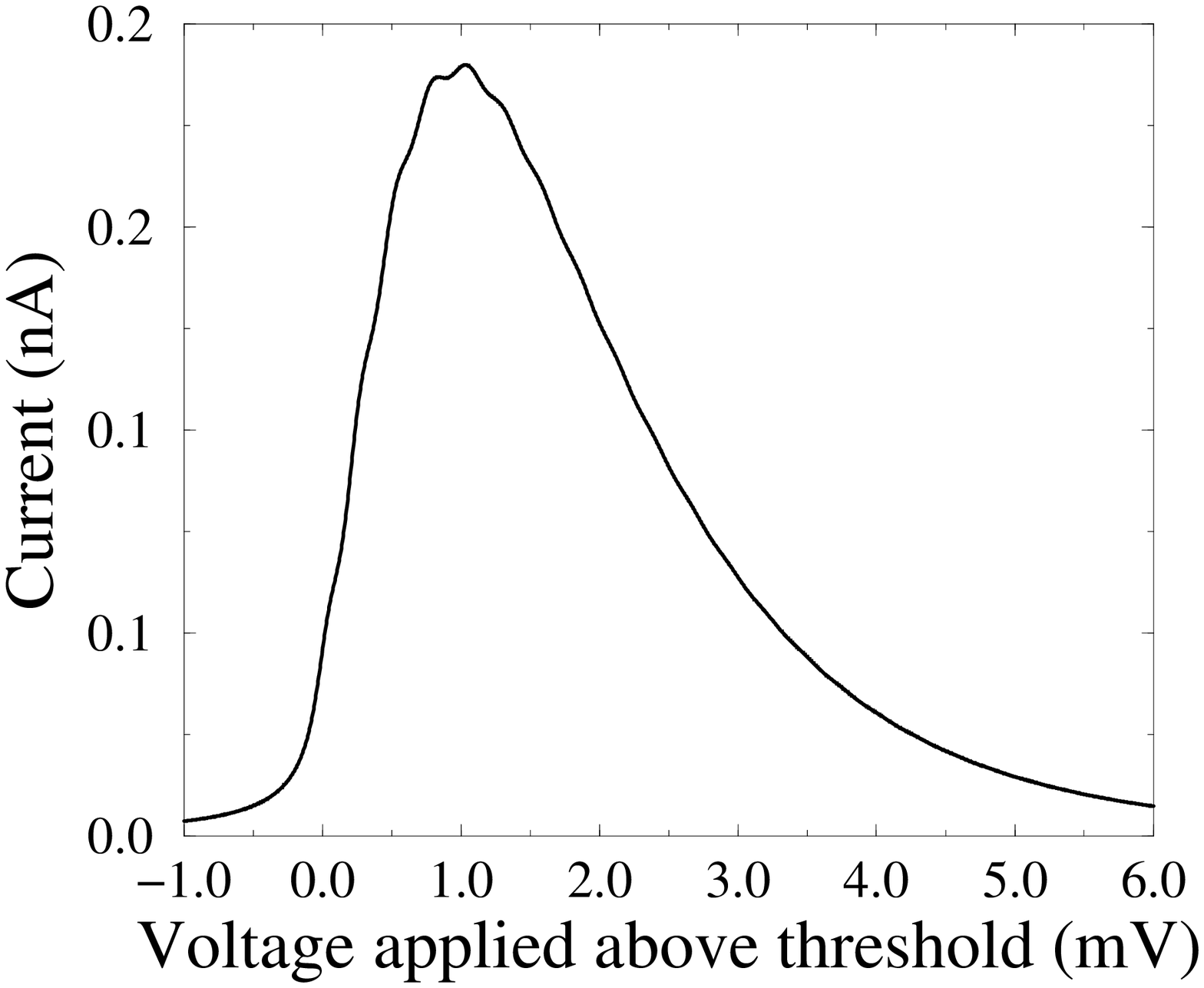,width=0.7\linewidth,angle=0}
\caption{Current at \(\nu=7/15\) as a function of \(V-V_0\). Finite
broadening as in Fig.~\ref{fig:CURRENT1}.}
\label{fig:CURRENT4}
\end{figure}

Figures~\ref{fig:CURRENT2},~\ref{fig:CURRENT4}~and~\ref{fig:CURRENT4} show the I-V characteristic
for the 2DEG at three odd-denominator filling factors originating from the
\(\nu=1/2\) state. A state
at \(\nu=p/(2mp+1)\) can be described in terms of composite fermions at
effective filling
factor \(\nu^\star=p\) interacting with flux tubes carrying \(2m\)
magnetic flux quanta. We have considered the cases where \(m=1\) and \(p=1,\ 3,\)
and 7, which 
correspond to the fractions \(\nu=1/3,\ 3/7\) and 7/15 respectively. 
The I-V characteristic at \(\nu=p/(2mp+1)\) differs in two important respects from that at 
\(\nu=1/2\). First, there is now no orthogonality catastrophe, because there is a gap to
excitations. While in the absence of the 2DEG the threshold current is \(e\Gamma_0\), 
the presence of the 2DEG reduces the threshold current by a factor of \(\exp(-L)\),
where  
\begin{equation}
L=\int_0^\infty d\omega\,\frac{\rho(\omega)}{\omega^2}
\end{equation}
is the Debye-Waller factor.
This current decreases as the effective magnetic field is reduced because \(L\)
increases, signifying the
reduction in the overlap of the initial state and final ground state; in the
limit of zero effective field \(L\rightarrow \infty\), implying that the overlap tends to 
zero yielding the OC. 
Second, the I-V spectrum develops oscillations 
superimposed on an envelope similar in shape to the I-V curve for
\(\nu=1/2\). 
The period of these oscillations is given by the composite fermion cyclotron
frequency \(\omega^\star_{\text{c}}=eB^\star/m^\star_{\text{CF}}\),
where \(B^\star=n_{\text{e}} h/e p\).
A measurement of this period can thus provide an experimental test of the  
validity of Eq.~(\ref{eq:MASS}).
\par
\begin{figure}
\centering\epsfig{file=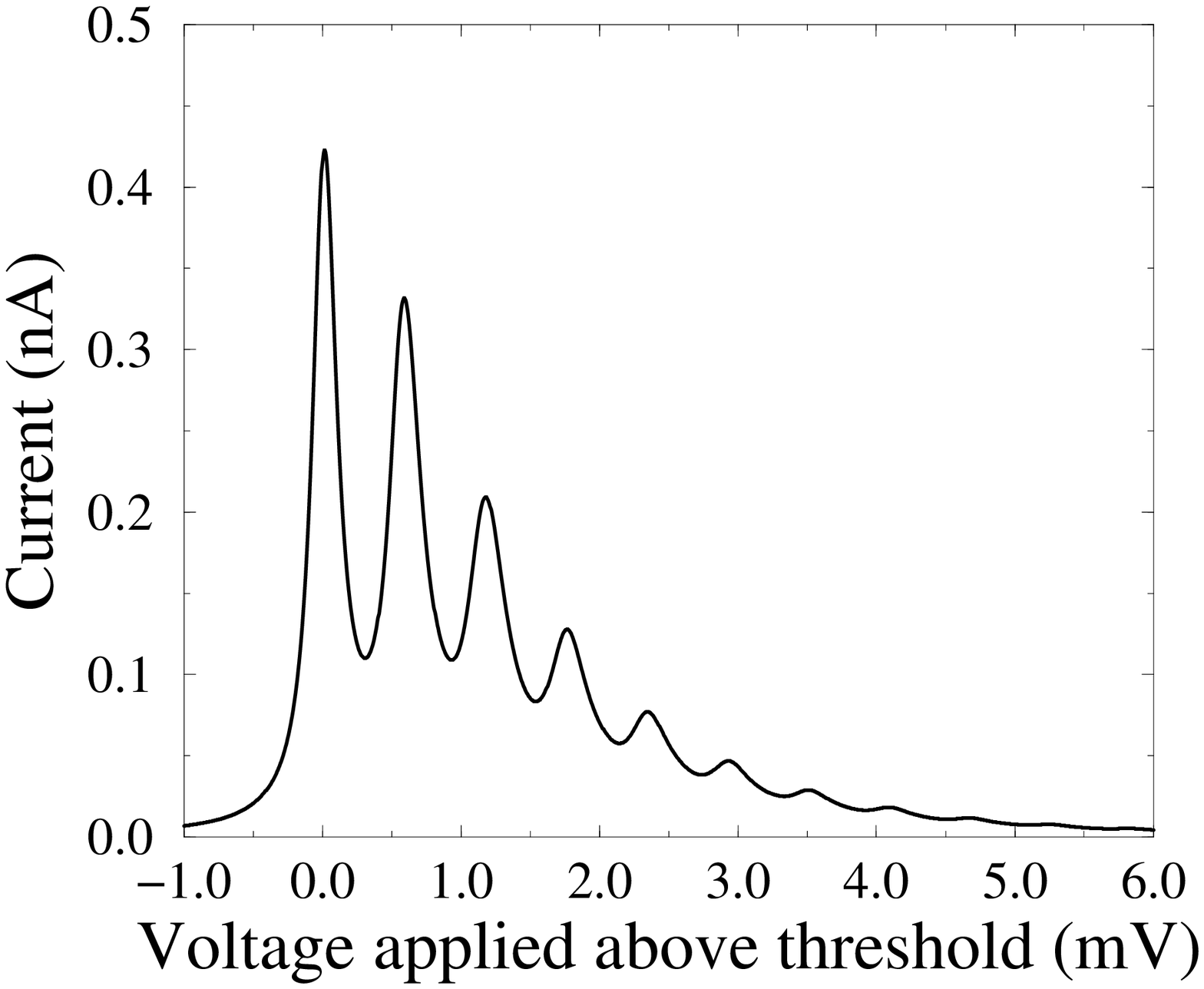,width=0.7\linewidth,angle=0}
\caption{Current at \(\nu=3/7\) as a function of \(V-V_0\). Finite
broadening as in Fig.~\ref{fig:CURRENT1}.}
\label{fig:CURRENT3}
\end{figure}

At high effective fields \(B^\star\), such as \(\nu^\star=1\) \((\nu=1/3\)),
the oscillations dominate the spectrum 
and the threshold current is large because \(L\) is small.
As \(B^\star\) is reduced \(L\) increases, thus the threshold current decreases
and the envelope function becomes more distinct. 
As expected the \(B^\star\rightarrow 0\) limit resembles the \(B^\star=0\)
calculation giving further confidence in our results.
The transition from high to low effective field displayed in Figs.~\ref{fig:CURRENT1}--~\ref{fig:CURRENT4}
is quite dramatic -- the peak at threshold changes to a complete suppression.
We now mention the shape of \(\rho(\omega)\) away from \(\nu=1/2m\). 
\(\rho(\omega)\) reflects the gapfull nature of
the excitations, containing peaks separated by \(\omega^\star_{\text{c}}\).
When \(B^\star\) is small, \(\rho(\omega)\) has an envelope which goes roughly as
\(\sqrt{\omega}\) for low energies and hence tends towards the \(B^\star=0\) (i.e. \(\nu=1/2m\))
shape.
\par
We now compare the response of an ordinary metal in a weak magnetic field
with the 
composite fermion system described above~\cite{MAHAN,SHAM}. The I-V
spectrum for ordinary  
electrons in zero magnetic field diverges as a power-law, being identically
zero at threshold. On applying a magnetic field the spectrum develops
oscillations with a period
equal to the cyclotron frequency but retains the divergent shape as an
envelope.
Since there is now a gap to excitations, there is no OC and \(I(V_0)=e\Gamma_0\,\exp(-L)\). 
However, even in the high field limit the power-law shape is retained in the envelope of the
oscillations. Hence the only changes in the overall shape of the I-V curve as the field is 
increased are the appearance of oscillations and the disappearance of the 
orthogonality catastrophe. This is in sharp contrast to the present case of
composite fermions where, as
shown above, the near-threshold behaviour changes from being completely
suppressed in zero effective field (e.g. \(\nu=1/2\)), to
being highly peaked, at large effective fields \((\text{e.g. }\nu^\star=1, \text{ and hence }
\nu=1/3)\).
\par
We thank B. I. Halperin and S. H. Simon for useful discussions. 
We acknowledge the financial support of EPSRC through a Studentship (D.J.T.L) and
EPSRC grant No. GR/K 15619.

\end{document}